\documentclass{article}
\usepackage[T1]{fontenc} % add special characters (e.g., umlaute)
\usepackage[utf8]{inputenc} % set utf-8 as default input encoding
\usepackage{ismir,amsmath,cite,url}
\usepackage{graphicx}
\usepackage{color}

\usepackage[leftmargin=1.2em,rightmargin=1.2em,vskip=0.3em,font=itshape]{quoting}

\title{\proposed: An interface for exploring\\ both text prompts and audio priors\\ in generating music with text-to-audio models}

\twoauthors
 {Hiromu Yakura} {University of Tsukuba \\ Tsukuba, Japan \\ {\tt hiromu.yakura@aist.go.jp}}
 {Masataka Goto} {National Institute of Advanced Industrial Science \\ and Technology (AIST), Tsukuba, Japan \\ {\tt m.goto@aist.go.jp}}

\def\authorname{H. Yakura and M. Goto}

\usepackage[bookmarks=false,pdfauthor={\authorname},pdfsubject={\papersubject},hidelinks]{hyperref}
\newcommand{\ie}{\textit{i.e.},~}
\newcommand{\etal}{\textit{et al.}~}

\newcommand{\proposed}{IteraTTA}

\makeatletter
\g@addto@macro\@author{
    \vspace{0.5cm}
    \\
    \hspace{-0.8cm}
    \makebox[0pt]{
        \includegraphics[width=0.9\linewidth]{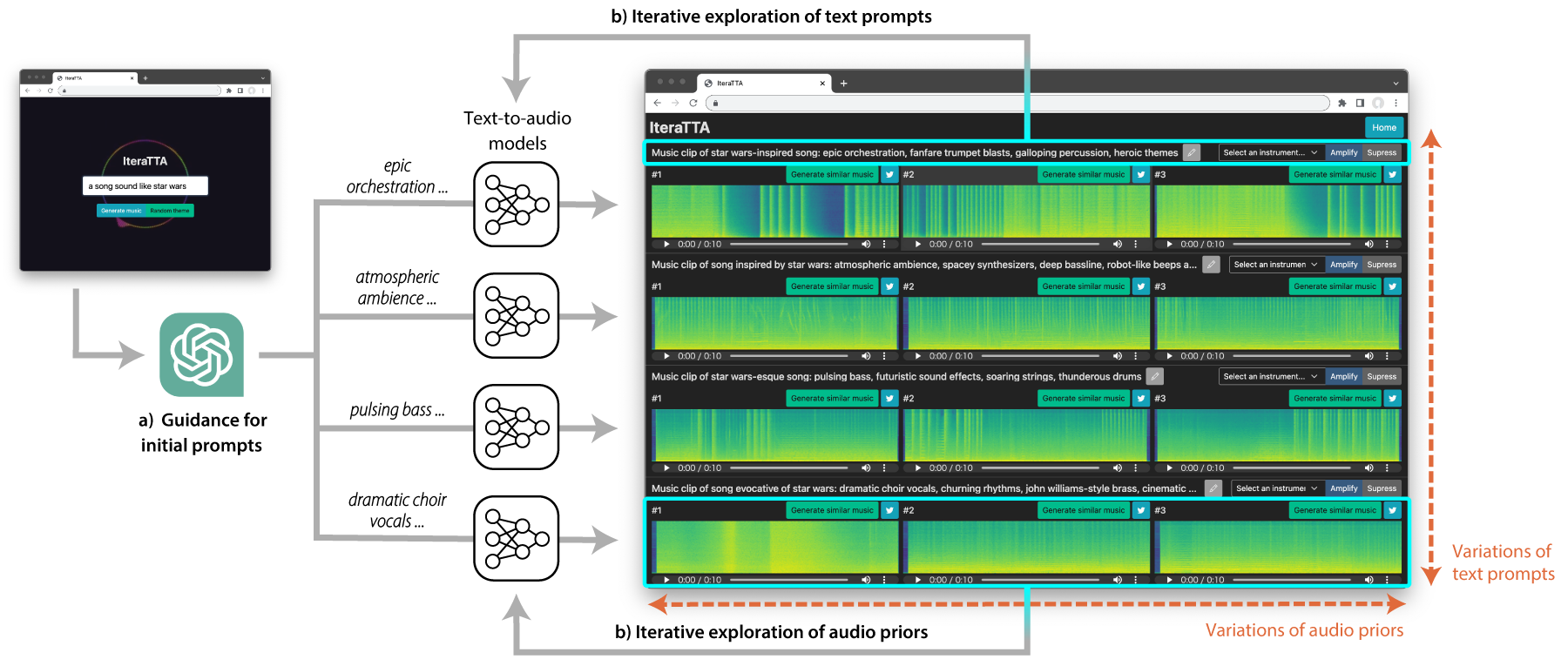}
    }
    \vspace{0.2cm}
    \\
    \parbox{0.97\linewidth}{
        \refstepcounter{figure}
        \normalsize
        \textbf{Figure~\thefigure}.
        \proposed{} is an interface dedicated for allowing novice users to show their creativity in text-to-audio music generation processes.
        It provides a) computational guidance for constructing initial prompts and b) dual-sided iterative exploration of text prompts and audio priors.
        \label{fig:hero}
    }
    \hfill
    \vspace{-0.5cm}
}
\makeatother

\begin{document}

\maketitle

\begin{abstract}
Recent text-to-audio generation techniques have the potential to allow novice users to freely generate music audio.
Even if they do not have musical knowledge, such as about chord progressions and instruments, users can try various text prompts to generate audio.
However, compared to the image domain, gaining a clear understanding of the space of possible music audios is difficult because users cannot listen to the variations of the generated audios simultaneously.
We therefore facilitate users in exploring not only text prompts but also audio priors that constrain the text-to-audio music generation process.
This dual-sided exploration enables users to discern the impact of different text prompts and audio priors on the generation results through iterative comparison of them.
Our developed interface, \proposed{}, is specifically designed to aid users in refining text prompts and selecting favorable audio priors from the generated audios.
With this, users can progressively reach their loosely-specified goals while understanding and exploring the space of possible results.
Our implementation and discussions highlight design considerations that are specifically required for text-to-audio models and how interaction techniques can contribute to their effectiveness.
\end{abstract}

\section{Introduction}
\label{sec:introduction}

Recent advances in generative machine learning techniques open up novel ways for a diverse group of individuals to engage in creative processes~\cite{DBLP:journals/corr/abs-2104-02726,DBLP:conf/chi/MullerCKMW22}.
Specifically, music generation models can foster creative expression among novice users, who may not necessarily possess formal musical knowledge~\cite{DBLP:journals/frai/CarnovaliniR20,DBLP:journals/tismir/Rohrmeier22}.
Consequently, several approaches have been proposed to enable users to control various musical attributes of generated audios, such as specifying the note or rhythm density~\cite{DBLP:conf/ismir/TanH20,DBLP:conf/ismir/0021WBD20} and chord progression~\cite{DBLP:conf/ismir/0008WZX20,DBLP:conf/ismir/DaiJGD21,DBLP:conf/ismir/0008X21}.
Text-to-audio models~\cite{DBLP:journals/corr/abs-2301-11325,DBLP:journals/corr/abs-2301-12503} are promising in terms of allowing users who are not familiar with the concepts of such musical attributes to generate their own sounds.

Nevertheless, there are still several gaps toward deploying such models to support the creativity of novice users.
For example, the models rely on annotated labels of music clips presented in their training datasets~\cite{DBLP:conf/icassp/GemmekeEFJLMPR17,DBLP:conf/naacl/KimKLK19,DBLP:conf/icassp/DrossosLV20}, which primarily consist of musical descriptions such as genres, instruments, and moods.
Therefore, providing such information as a text prompt is crucial for enabling fine-grained control over generated music audios.
However, this may prove challenging for novice users due to disparities in artistic vocabulary among individuals with varying levels of musical knowledge~\cite{Swanwick2002}.
Experimentally, it has been suggested that non-musicians tend to rely more on abstract concepts, such as the pleasantness or complexity of music, when appreciating musical pieces~\cite{Gromko1993}, which may pose difficulties in fully exploring various text prompts.

Moreover, understanding the space of possible results is also challenging, particularly when compared to the use of text-to-image models.
In text-to-image generation, users can look over various generation results at a glance, which fosters their understanding of the space and helps them decide on directions to explore~\cite{DBLP:journals/corr/abs-2303-13534}.
From the perspective of explainable AI (XAI), we can say that such results serve as \textit{explanations by example} \cite{DBLP:journals/inffus/ArrietaRSBTBGGM20} because the results implicitly invite the users to infer the behavior of the models.
However, in text-to-audio generation, users cannot simultaneously listen to multiple generation results, thus impeding their comprehension and ability to efficiently explore the space.
These points imply that specific design considerations are necessary to fully leverage the potential of text-to-audio models and exploring them would also provide a new perspective in terms of XAI.

In this paper, we introduce \proposed{}, an interface dedicated to the text-to-audio (TTA) music generation processes of novice users.
This interface enables iterative exploration of both text prompts and audio priors, allowing users to gain a comprehensive understanding of the space of possible results by sufficiently constraining the generation processes.
We constructed this interface based on our observations and related literature on creativity support, which emphasize the importance of 1) computational guidance for constructing initial prompts and 2) dual-sided iterative exploration of text prompts and audio priors.
Moreover, we deployed the interface as a publicly-available Web service and analyzed the diverse ways in which users utilized it in their creative processes.
Our results and discussions shed light on ways to utilize models developed in the MIR community to unleash the creativity not only of expert users~\cite{DBLP:conf/ismir/AndersenK16} but also of individuals with varying degrees of musical knowledge.

\section{Related Work}
\label{sec:rw}

\subsection{Music Generation Techniques}

Music generation has been one of the central topics with the MIR community~\cite{DBLP:journals/csur/Roads85,DBLP:journals/jair/FernandezV13,DBLP:journals/tetci/LiuT17,DBLP:journals/nca/BriotP20,DBLP:journals/tismir/DerutyGLNA22}, and recently, generative machine learning techniques have been widely employed for this purpose~\cite{DBLP:journals/corr/abs-2011-06801,DBLP:journals/tismir/DerutyGLNA22}.
While methods for symbolic music generation that output MIDI files have been popular~\cite{DBLP:conf/ismir/YangCY17,DBLP:conf/ismir/DongY18,DBLP:conf/iclr/HuangVUSHSDHDE19,DBLP:conf/mm/HuangY20,DBLP:conf/aaai/HsiaoLYY21,DBLP:conf/ismir/MittalEHS21,DBLP:journals/corr/abs-2210-10349}, some methods use generative models to directly output audio, leveraging their expressiveness~\cite{DBLP:journals/corr/abs-2005-00341,DBLP:journals/corr/abs-2111-05011,DBLP:conf/ismir/HungCYY21,DBLP:journals/corr/abs-2208-08706}.
For example, Jukebox~\cite{DBLP:journals/corr/abs-2005-00341} and RAVE~\cite{DBLP:journals/corr/abs-2111-05011} use variational autoencoders and autoregressive models trained on large-scale music datasets to generate diverse music audios.

Controllability in music generation has been also emphasized~\cite{DBLP:conf/ismir/TanH20,DBLP:conf/ismir/0021WBD20,DBLP:conf/ismir/0008WZX20,DBLP:conf/ismir/DaiJGD21,DBLP:conf/ismir/0008X21,DBLP:conf/ismir/Akama20,DBLP:conf/ismir/HungCDKNY21,DBLP:journals/corr/abs-2212-11134} because it is vital to open up its applications for supporting users' creative processes~\cite{DBLP:journals/chb/WangN17,DBLP:conf/ismir/HuangKNDC20}.
For instance, Music FaderNets~\cite{DBLP:conf/ismir/TanH20} allows users to modify the rhythm and note densities of generation results, while Music SketchNet~\cite{DBLP:conf/ismir/0021WBD20} enables them to specify pitch contours and rhythm patterns.
Wang \etal{}\cite{DBLP:conf/ismir/0008WZX20} and Dai \etal{}\cite{DBLP:conf/ismir/DaiJGD21} have proposed methods to further constrain the chord progression of generation results.
However, as mentioned in \secref{sec:introduction}, users are not always familiar with such concepts, and then, they would have difficulties in using these methods to output music audios they want to generate.
We acknowledge that some methods~\cite{DBLP:conf/ismir/HungCDKNY21,DBLP:journals/corr/abs-2212-11134} provide perceptual control that does not require extensive musical knowledge: emotion-based musical generation.
Nevertheless, they are based on Russell’s valence-arousal model~\cite{Russell1980} consisting of four classes, which limits the range of controls and may hamper users' agency~\cite{Heer2019} when the methods are used to support their creative processes.

In this context, recent text-to-audio models~\cite{DBLP:journals/corr/abs-2301-11325,DBLP:journals/corr/abs-2301-12503} can be an effective tool for such novice users.
These models learn the relationship between music audios and their text descriptions (more specifically, latent representations encoded from the descriptions by RoBERTa~\cite{DBLP:journals/corr/abs-1907-11692}) and use it to guide results in generating new audios from an inputted text (\ie{}text prompt).
As RoBERTa can encode text prompts with variable length and content, the models can provide flexible control without requiring specific musical knowledge of rhythm patterns or chord progressions.
Moreover, they allow users to constrain generation results not only by text prompts but also by audio priors, ensuring that the results have similar characteristics to the priors.
For example, the diffusion model~\cite{DBLP:conf/cvpr/RombachBLEO22} employed by AudioLDM~\cite{DBLP:journals/corr/abs-2301-12503} usually uses Gaussian noise for the seed of its generation process, but by using a noise-infused audio prior, we can obtain generation results preserving the characteristics of the provided audio.

Here, text-to-image models that use similar schemes have been shown to unleash the creativity of novice users, allowing them to iteratively explore open-ended variations of text prompts~\cite{DBLP:journals/corr/abs-2303-13534} and customize their intermediate results by specifying image prior constraints~\cite{DBLP:conf/eccv/GafniPASPT22}.
Similarly, text-to-audio models can be leveraged to provide users with such iterative exploration or customization.
However, we also expect that text-to-audio music generation processes may pose several specific difficulties, as explained in \secref{sec:introduction}.
Therefore, we explored how interaction techniques can address these challenges by developing an interface dedicated to text-to-audio models.

\subsection{Interfaces for Music Generation}

There is a series of research on building interfaces to let users interact with music generation techniques effectively~\cite{DBLP:conf/chi/SimonMB08,DBLP:journals/corr/abs-1907-06637,DBLP:conf/chi/LouieCHTC20,simeon_rau_2022_7316618,DBLP:conf/nime/ZhangXLD21,DBLP:journals/corr/abs-2010-03190,DBLP:conf/iui/Zhou0GI21}.
MySong~\cite{DBLP:conf/chi/SimonMB08}, for instance, involves a music accompaniment generation model, with which users can control the happiness or jazziness of generation results.
Louie \etal{}\cite{DBLP:conf/chi/LouieCHTC20} proposed an interactive interface for novice users so that they can use a symbolic music generation technique with control of happiness or randomness.
The interface also allows users to constrain generation results by providing music priors, which was experimentally confirmed to be effective in iteratively refining the results.
Zhou \etal{}\cite{DBLP:journals/corr/abs-2010-03190,DBLP:conf/iui/Zhou0GI21} utilized a user-in-the-loop Bayesian optimization technique to enable novice users to iteratively explore melodies composed by a generative model.

These interfaces underscore the significance of providing controls and supporting iterative exploration in facilitating the creativity of novice users using music generation techniques.
Consequently, the provision of recent text-to-audio models to novice users would be highly suitable for this purpose, as they offer more flexible control, compared to using several parameters such as happiness, while also allowing the use of audio priors.
Our paper contributes to this series of research by examining design considerations of interfaces for text-to-audio music generation processes, aiming to expand the scope of applications of recent techniques developed in the MIR community.

\section{Design Requirements}
\label{sec:requirements}

As stated in \secref{sec:introduction}, our goal is to leverage text-to-audio models to facilitate the creative expression of novice users regardless of their musical knowledge.
To this aim, we embarked upon an examination of potential challenges that these users may encounter during text-to-audio music generation processes and subsequently derived a set of design requirements to address these issues.
Guided by the principles of human-computer interaction, we utilized the think-aloud protocol~\cite{DBLP:journals/sigchi/WrightM91,DBLP:conf/chi/AlhadretiM18} by involving three volunteers who self-reported that they possessed no formal musical training beyond compulsory education.
Specifically, we provided the volunteers with access to one of the latest text-to-audio models~\cite{DBLP:journals/corr/abs-2301-12503} on Google Colab using its official implementation\footnote{\url{https://github.com/haoheliu/AudioLDM}}, which enabled them to provide any text prompts and subsequently listen to three music audios generated from the text prompts.
Here, since the remotely-participated volunteers were Japanese speakers recruited via word-of-mouth communication, we told them that they can use DeepL Translator to translate text prompts into English to obtain better results with the model that is mainly trained on the dataset with English text labels~\cite{DBLP:conf/icassp/GemmekeEFJLMPR17,DBLP:conf/naacl/KimKLK19,DBLP:conf/icassp/DrossosLV20}.
They freely used the model for approximately 30 minutes while sharing their screens on a video call and verbalizing their thoughts and feelings.
This allowed us to identify the challenges that they encountered and the factors that contributed to these challenges.
We then conducted semi-structured interviews to validate the challenges identified and to gain further insight into the reasons behind them.
Their responses were analyzed based on open coding \cite{StrCor90}, which yielded the following design requirements in line with the existing literature on creativity support.

\subsection{Computational guidance for constructing initial prompts}
\label{sec:requirements-gpt}

We observed that the volunteers frequently encountered difficulty in formulating appropriate text prompts to initiate their use of the model.
For example, one volunteer entered the phrase ``a song sounds like star wars,'' resulting in audio containing a battle cry with a space-like sound effect.
This can be attributed to the characteristics of the text labels in the dataset used to train the model~\cite{DBLP:conf/icassp/GemmekeEFJLMPR17,DBLP:conf/naacl/KimKLK19,DBLP:conf/icassp/DrossosLV20}.
Specifically, the labels of music clips consist primarily of musical descriptions such as genres, instruments, and moods, like: ``An orchestra plays a happy melody while the strings and wind instruments are being played~\cite{DBLP:conf/icassp/DrossosLV20}.''
Therefore, providing such a description would be essential to ensure that the model trained on the dataset generates music audio as intended.
The volunteer was unable to generate music-like audio until he attempted several prompts and finally entered ``solemn music starting with a trumpet fanfare.''

In the context of creativity support, two underlying factors could explain the aforementioned observation.
First, an inherent gap in artistic vocabulary exists between expert and novice users~\cite{Swanwick2002}.
Without deep musical knowledge, it can be challenging to conceive a precise description of music audios.
Additionally, novice users often have loosely-specified goals when starting a creative endeavor~\cite{DBLP:journals/tog/TaltonGYHK09,10.5555/1891970.1891971,DBLP:conf/ijcai/Yakura0G21}.
They refine their objectives gradually by exploring the space of possible results through iterative exploration~\cite{DBLP:conf/bcshci/Tweedie95,DBLP:conf/uist/TerryM02}.
However, the dependency of text-to-audio models on precise descriptions of clearly-defined goals makes it difficult for novice users to initiate such exploration.
This suggests that supporting them computationally in constructing initial prompts could potentially facilitate the creativity of novice users.

\subsection{Dual-sided iterative exploration of text prompts and audio priors}
\label{sec:requirements-dual}

\begin{figure*}[t]
    \centering
    \includegraphics[width=0.7\textwidth]{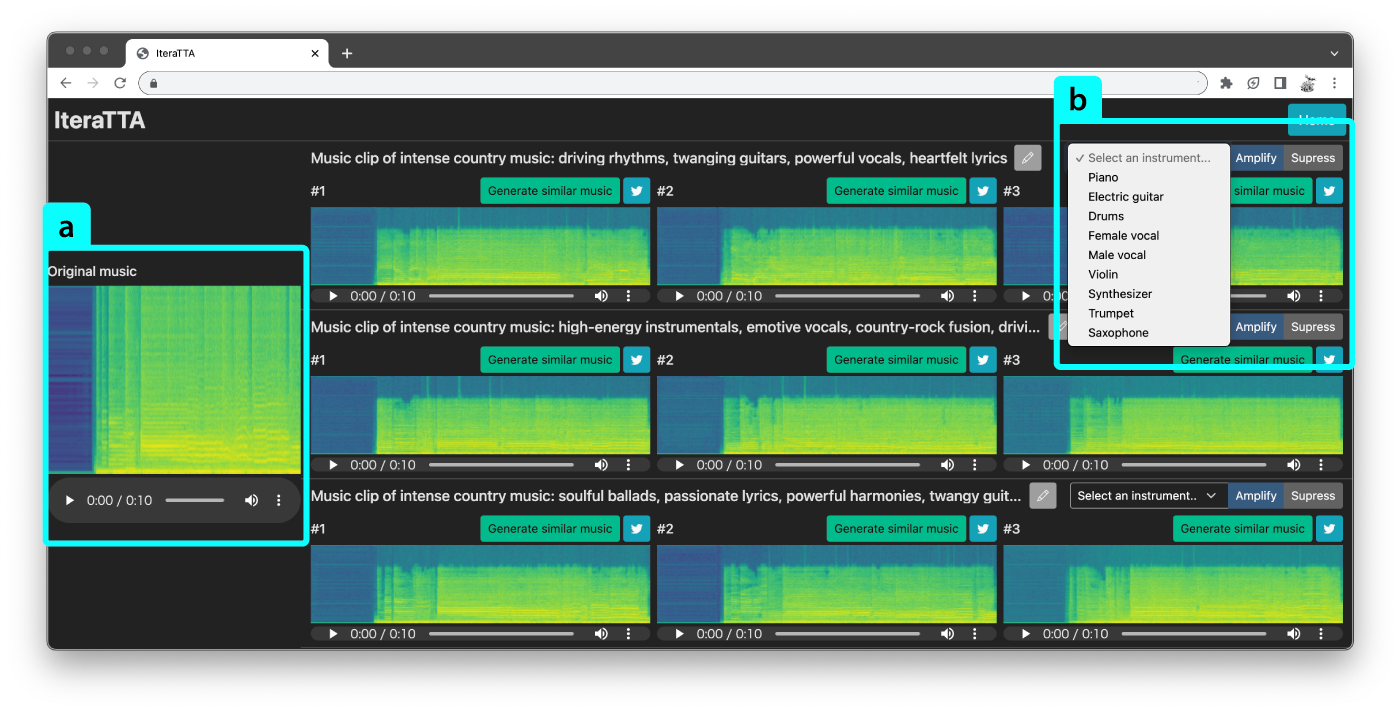}
    \caption{To facilitate the exploration of text prompts and audio priors, \proposed{} allows a) comparison of generation results with an audio prior and b) instant edit of a text prompt.}
    \label{fig:feature}
\end{figure*}

We also observed that the volunteers encountered challenges in efficiently exploring the generated results.
One volunteer who had prior experience with text-to-image models mentioned the point, as:
\begin{quoting}
    ``Unlike text-to-image models, comparing various results at a glance was difficult with the text-to-audio model. So, finding a text prompt reflecting my intention most faithfully became much tough.''
\end{quoting}
In other words, iteratively trying different text prompts would not necessarily assist users in comprehending the space of potential results, although it is necessary for novice users to refine their loosely-specified goals~\cite{DBLP:conf/bcshci/Tweedie95,DBLP:conf/uist/TerryM02}.
Therefore, users cannot determine which direction would be closest to their goals and what text prompt to try next.
Another volunteer mentioned an issue he faced, as:
\begin{quoting}
    ``I once found a generation result with a good melody, but I wanted to change its tone. So, I added `with a flute' to its text prompt and regenerated. However, the melody was then completely changed, which was frustrating.''
\end{quoting}
This implies that we need to let users utilize not only text prompts but also audio priors to constrain the tune of generation results.
In sum, supporting the creativity of novice users in text-to-audio music generation processes requires enabling them to efficiently explore variations of both text prompts and audio priors, allowing them to iteratively refine their goals by understanding the space of possible results.
This demands us to develop an interface specifically tailored for text-to-audio models to provide such dual-sided exploration of text prompts and audio priors.

\section{\proposed{}}

Based on the above design requirements, we present \proposed{}, a dedicated interface for text-to-audio music generation processes.
It was implemented as a Web-based system, allowing novice users to instantly benefit from the latest text-to-audio models in their creative processes.

\subsection{Design}

As illustrated in \figref{fig:hero}, our interface requires users to first input a theme phrase for music audios to generate.
The inputted phrase need not include precise musical descriptions since \proposed{} leverages a large language model to derive such descriptions suitable for text-to-audio models using knowledge embedded in the models~\cite{DBLP:journals/corr/abs-2206-07682}.
Specifically, the interface queries a large language model that ``Please give me four variational lists of comma-separated phrases describing what does a music clip of "\textit{[theme phrase]}" sound.''
It then uses the four responded phrase lists as a variety of the first text prompts to start the music generation processes in parallel.
This feature allows novice users to translate loosely-specified goals in their minds into musical descriptions, which can also help them to envisage variations of text prompts to explore.

\proposed{} then generates three music audios for each of the four prompts.
The generated audios are arranged in two dimensions (see \figref{fig:hero}), which enables novice users to understand how different music audios are generated by different text prompts, and also, how different music audios are generated by the same text prompts.
This is intended to assist users in identifying which text prompts and audio priors are closely aligned with their goals and which direction is worth exploring.
If a user identifies a suitable candidate text prompt, they can customize the prompt and generate new music audios with it.
Alternatively, if the user discovers a suitable music audio, they can use it as an audio prior to generate new music audios.
In essence, the user can explore the subspace of possible results that are proximate to their goals by constraining either text prompts or audio priors, while gradually refining their goals by themselves.

We have incorporated several features to facilitate the exploration of text prompts and audio priors, as shown in \figref{fig:feature}.
For instance, when a user specifies an audio prior, \proposed{} enables the user to compare generated results with it.
It also offers an instant editing feature of text prompts, allowing users to amplify or suppress the sound of a selected instrument.
This is achieved by simply adding a phrase of "with strong \textit{[instrument]}" or "with no \textit{[instrument]}" into a text prompt, but it provides an example of how they can modify generation results through prompts.

\subsection{Implementation}

As mentioned, we developed \proposed{} as a Web-based system to invite novice users for trying music generation with it.
For the implementation of its back-end server, we utilized Python with FastAPI and incorporated an API of GPT-3.5\footnote{We used \texttt{gpt-3.5-turbo} of \url{https://platform.openai.com/docs/models/gpt-3-5}.} to construct initial prompts, while AudioLDM~\cite{DBLP:journals/corr/abs-2301-12503} was employed to generate the music audios.
The length of music audios to generate was predetermined at 10 seconds so that our GPU server harnessing an NVIDIA RTX 2080 Ti can afford the generation of 12 audios (3 audios $\times$ 4 prompts) simultaneously.
On average, the generation process takes approximately 15 seconds.
In addition, we used DeepL API to translate text prompts into English when they were provided in non-English languages because we observed that it led to better results in \secref{sec:requirements}.
For the front-end interface of \proposed{}, we utilized Vue.js, which enables users to download the generated music audios or share them on Twitter.

\begin{figure*}[t]
    \centering
    \includegraphics[width=0.83\textwidth]{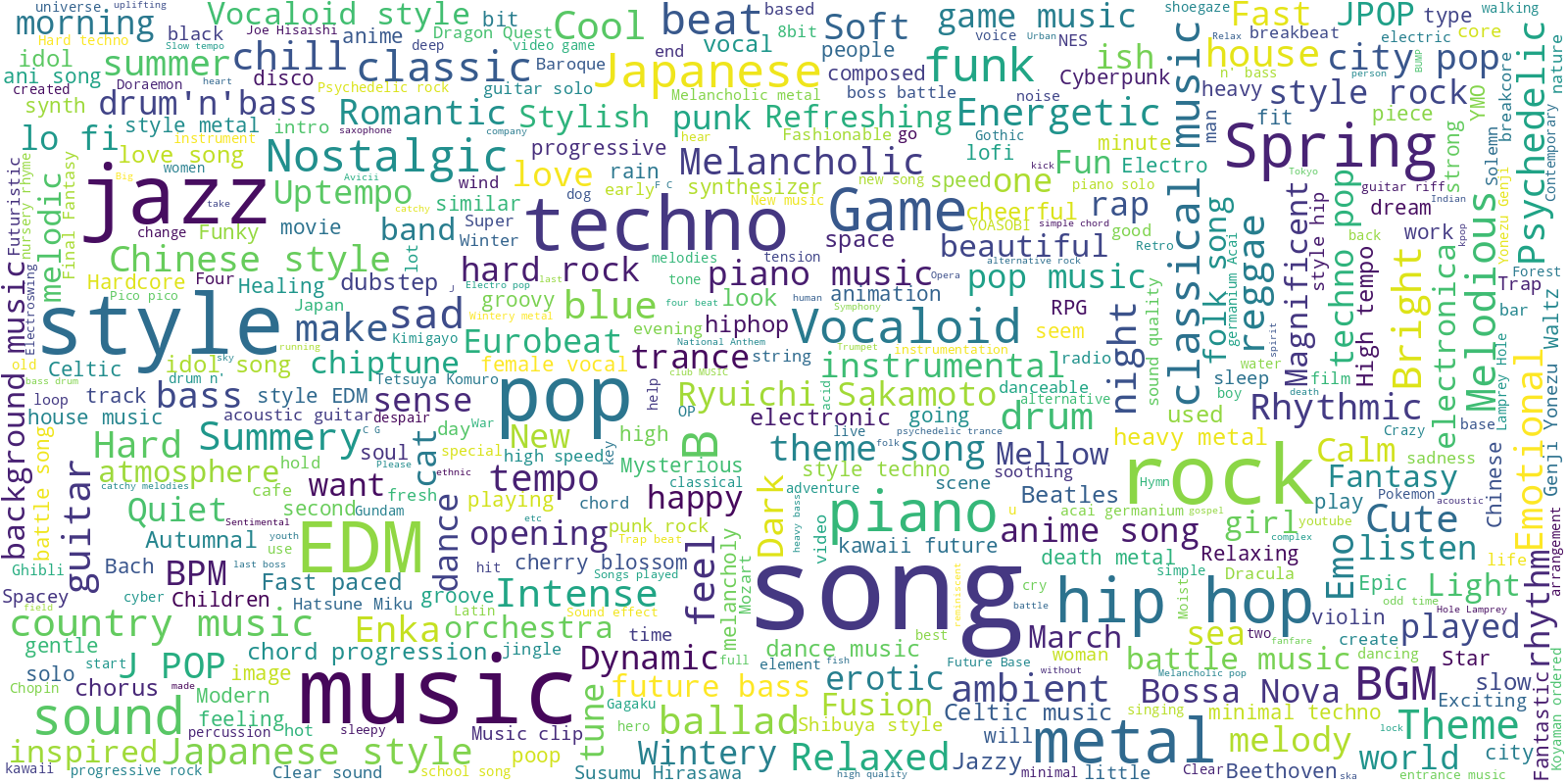}
    \caption{Word cloud of theme phrases the 8,831 users inputted on our Web service.}
    \label{fig:wordcloud}
\end{figure*}

\section{Analysis}

To investigate the effectiveness of \proposed{} in supporting diverse users in the wild, we deployed it as a publicly-available Web service in Japanese\footnote{Its English version is currently available at \texttt{http://iteratta.duckdns.org/}, and readers can try it on their Web browsers (Google Chrome is recommended).}.
Within two days of release, 8,831 users generated 246,423 music audios.
In this section, we discuss the insights we extracted from their usage logs and their responses to a form that we put a link to it on the Web service so that they can share their opinion and feedback voluntarily.

\subsection{Diversity of theme phrases}

We first examined the theme phrases that users inputted to initiate text-to-audio music generation processes and found that they were highly varied.
Some users provided music-related phrases, such as ``nice city pop'' and ``cute future bass,'' while others were more specific, like: ``80's hip hop that break dancers would dance to.''
There were also phrases expressing more abstract ideas, such as ``Arabian caves'' and ``silent dream of a priestess.''
\figref{fig:wordcloud} visualizes the words often used in the translated phrases in the form of a word cloud, showing their diversity.

\begin{figure}[t]
    \centering
    \includegraphics[width=0.97\columnwidth]{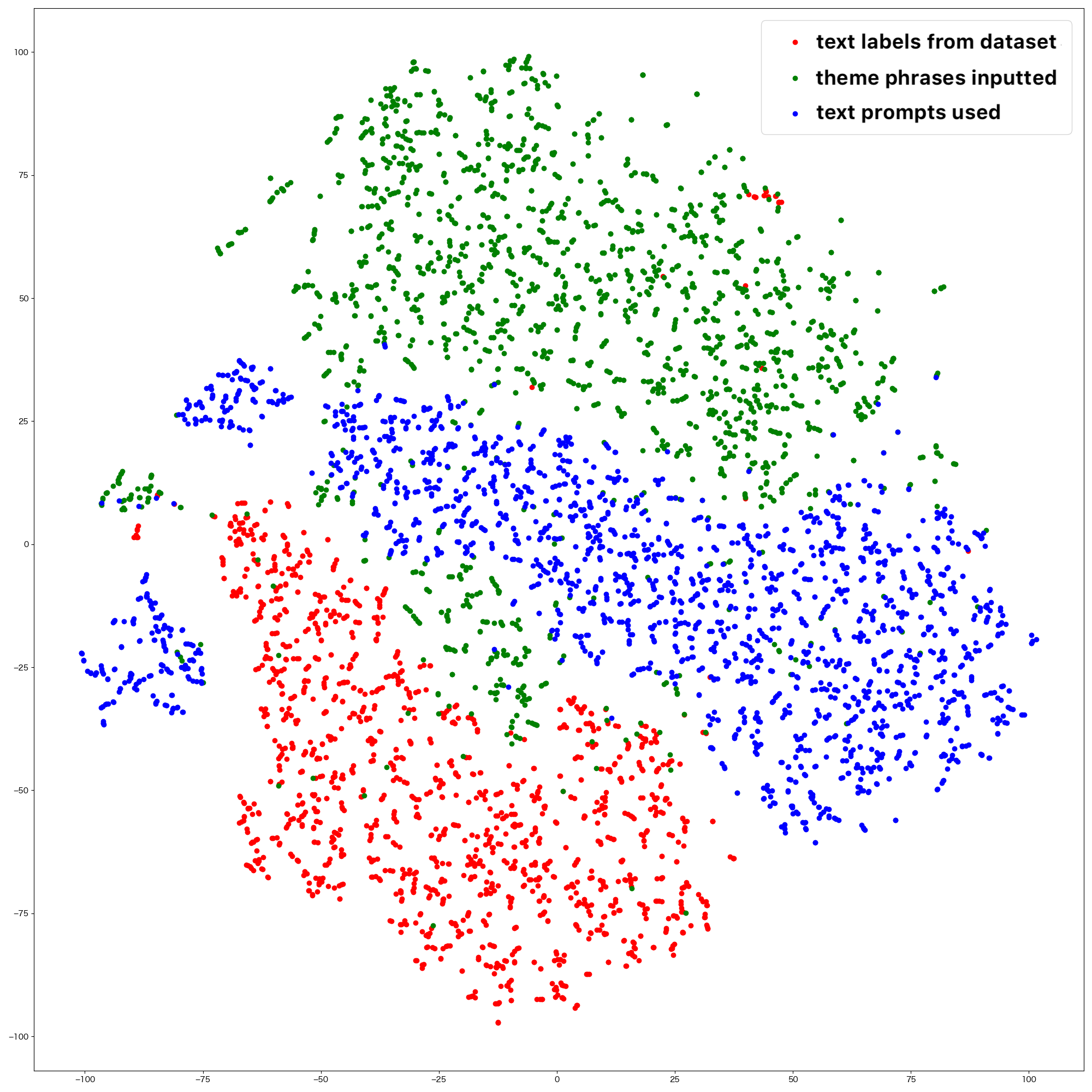}
    \caption{Visualization of the representation vectors of the theme phrases inputted by the users, the text prompts computationally derived from them, and the text labels in the training dataset.}
    \label{fig:tsne}
\end{figure}

To explore the role of \proposed{}, we compared the theme phrases inputted by the users and the text prompts derived from them by the large language model to the text labels in the dataset used for training the text-to-audio model.
Specifically, we randomly sampled 1,000 cases for each of the theme phrases, text prompts, and text labels\footnote{For the text labels, we extracted labels containing ``music'' from AudioCaps~\cite{DBLP:conf/naacl/KimKLK19}.} and calculated their representation vectors using the same pretrained model of RoBERTa~\cite{DBLP:journals/corr/abs-1907-11692} as the text-to-audio model.
We then visualized the distribution of the vectors using t-SNE~\cite{JMLR:v9:vandermaaten08a}, as presented in \figref{fig:tsne}.
This indicates that \proposed{} guided the large language model to derive text prompts that bridged the gap between the diverse users' theme phrases and the text labels in the training dataset.
In fact, we found that the large language model successfully derived text prompts containing musical descriptions even from abstract phrases, such as ``otherworldly harmonies, delicate strings, minimalistic percussion, dreamlike vocals'' for ``silent dream of a priestess.''
These results suggest the effectiveness of guiding the construction of initial prompts to support the creative processes of novice users, as discussed in \secref{sec:requirements-gpt}.

\subsection{Journey of iterative exploration}

\begin{figure*}[t]
    \centering
    \includegraphics[width=0.75\textwidth]{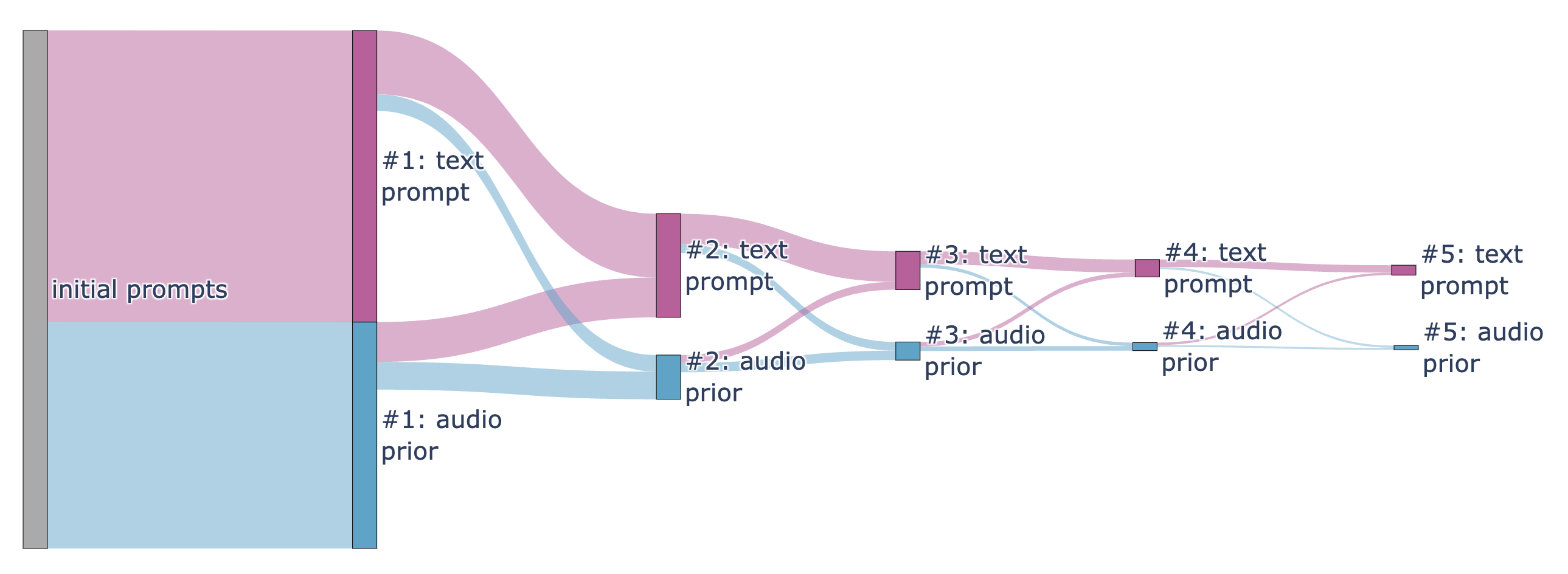}
    \caption{Visualization of how the users utilized the dual-sided exploration of \proposed{}.}
    \label{fig:sankey}
\end{figure*}

We also investigated how the users interacted with generated results produced by \proposed{}.
We analyzed the interaction log of the service and obtained \figref{fig:sankey}.
While some users just tried the exploration feature once, we found that others made iterative use of the feature, alternating between providing text prompts and audio priors.
Interestingly, one user repeated this refinement process 32 times, specifying text prompts 14 times and audio priors 18 times before sharing their final result on Twitter.
These points imply that our design, which enables dual-sided iterative exploration, helped the users effectively utilize the text-to-audio model.

\subsection{Unleashing the creativity of novice users}

We lastly analyzed the users' responses to the feedback form, which received 33 responses in total.
Overall, most of them expressed their affirmative experiences with the text-to-audio music creation processes, like:
\begin{quoting}
    ``It was a very interesting trial. I can interact with it throughout the day.''
\end{quoting}
\begin{quoting}
    ``In my personal opinion, it can be used as a source of sampling materials and an idea generator. As a person who usually composes music, I never had any negative feelings about composing from text using this. It is wonderful.''
\end{quoting}
The latter comment suggests that the features of \proposed{} prepared for novice users can also benefit experienced users in different ways.

It is also notable that the users left comments implying the importance of the design requirements discussed in \secref{sec:requirements}, such as how they enjoyed the open-ended exploration starting from loosely-specified theme phrases. 
\begin{quoting}
    ``It was fun to encounter songs that fit the theme I provided but I had never heard before.''
\end{quoting}
\begin{quoting}
    ``I really enjoyed the points that I could take advantage of ChatGPT's ability to associate and verbalize even seemingly unconnected ideas, which allowed me to provide crazy theme phrases that would not be understood by a human. I also learned a lot about how to describe songs by looking at the derived text prompts.''
\end{quoting}

Interestingly, in the form, some users left a successful prompt that they reached after exploration:
\begin{quoting}
    ``I would like to report that including a phrase of `simple progression' or limiting the number of tracks yielded stabilized music audios, like: `Ideal harmonious song: balanced instrumentation, band sound, simple chord progressions, rhythmic drum patterns, catchy pop melody, up to 12 tracks.'\,''
\end{quoting}
\begin{quoting}
    ``Adding `clear sound quality' produces less noisy audios.''
\end{quoting}
It is surprising that, even though we provided no explicit description of the behavior of text-to-audio models, the users were able to gain such knowledge by themselves through the iterative exploration with \proposed{}.
While such \textit{prompt modifiers} (also known as \textit{quality boosters})~\cite{DBLP:journals/corr/abs-2204-13988} that influence results in a specific way have been discovered for text-to-image models in a community-driven manner~\cite{DBLP:journals/corr/abs-2204-13988,DBLP:journals/corr/abs-2303-13534}, the above comments would be the first examples for text-to-audio models, to the best of our knowledge.
We assume that this is a manifestation of users' creativity in text-to-audio music generation processes and would be hard to derive without \proposed{}.

\section{Conclusion}

This paper introduces \proposed{}, an interface specifically designed for supporting novice users in their text-to-audio music generation processes.
Its design is guided by two main principles, providing a) computational guidance for constructing initial prompts and b) dual-sided iterative exploration of text prompts and audio priors.
The former can help novice users translate their loosely-specified goals into text prompts, which serve as starting points for exploration, even if they do not have rich artistic vocabularies.
The latter is important for enabling them to comprehend the space of possible results and gradually refine their goals.
To examine how diverse users utilize \proposed{} in their creative processes, we deployed it as a publicly-available Web service and analyzed users' behaviors, which highlight the importance of these design considerations in supporting the users' creativity.
Importantly, these principles are applicable not only to the specific text-to-audio model but to other models, including those to be proposed in the near future.
We believe that this paper can serve as a foundation for enabling novice users to benefit from state-of-the-art models in the MIR community.

\section{Acknowledgement}

This work was supported in part by JSPS KAKENHI Grant Number JP21J20353, JST ACT-X Grant Number JPMJAX200R, and JST CREST Grant Number JPMJCR20D4, Japan.

% For bibtex users:
\bibliography{reference}

\end{document}